\documentclass[11pt,twoside]{article}
\usepackage{asp2010}

\resetcounters

\begin{document}

\title{Creating White Dwarf Photospheres in the Laboratory: Strategy for Astrophysics Applications}
\author{Ross~E.~Falcon$^{1,2}$, G.~A.~Rochau$^2$, J.~E.~Bailey$^2$, J.~L.~Ellis$^1$, A.~L.~Carlson$^2$, T.~A.~Gomez$^1$, M.~H.~Montgomery$^1$, D.~E.~Winget$^1$, E.~Y.~Chen$^1$, M.~R.~Gomez$^2$, T.~J.~Nash$^2$, and T.~M.~Pille$^1$
\affil{$^1$Department of Astronomy and McDonald Observatory, University of Texas, Austin, TX 78712, USA}
\affil{$^2$Sandia National Laboratories, Albuquerque, NM 87185-1196, USA}}

\begin{abstract}
Astrophysics experiments by Falcon et al. to create white dwarf photospheres in the laboratory are currently underway.  The experimental platform measures Balmer line profiles of a radiation-driven, pure hydrogen plasma in emission and in absorption for conditions at $T_{\rm e}\sim 1$\,eV, $n_{\rm e}\sim 10^{17}$\,cm$^{-3}$.  These will be used to compare and test line broadening theories used in white dwarf atmosphere models.  The flexibility of the platform allows us to expand the direction of our experiments using other compositions.  We discuss future prospects such as exploring helium plasmas and carbon/oxygen plasmas relevant to the photospheres of DBs and hot DQs, respectively.
\end{abstract}

\section{Introduction}

In white dwarf (WD) astronomy, the method of infering photospheric conditions from the comparison of line profiles in observed spectra to those in synthetic spectra from WD atmosphere models, known as the spectroscopic method \citep[e.g.,][]{Bergeron92b}, is the most widely-used technique and is responsible for determining parameters for tens of thousands of WDs \citep[e.g.,][]{Liebert05,Kepler07,Koester09b,Castanheira10}.

This method is powerful and precise ($\frac{\delta T_{\rm eff}}{T_{\rm eff}}\sim5$\,\% and $\frac{\delta{\rm log}g}{{\rm log}g}\sim1$\,\% are typical for a given star), but it is not yet in its ultimate form.  The mean mass of WDs determined from spectroscopic investigations disagrees with the mean mass determined from the atmosphere model-independent technique which uses gravitational redshifts \citep{Falcon10}, indicating an underestimated spectroscopic mass at all $T_{\rm eff}$.  For cooler WDs with $T_{\rm eff}\lesssim12,000$\,K, fitting observed spectra is well-known to be particularly problematic \citep{Bergeron07,Koester09a}.  Furthermore, the WD atmosphere models used in the spectroscopic method are still advancing.  The latest models of \citet{Tremblay09} use newly-calculated Stark broadened line profiles of hydrogen that include non-ideal gas effects, and in re-analyzing the hydrogen-atmosphere WDs from the Palomar-Green Survey 
\citep{Liebert05}, their spectroscopic fits yield significant systematic increases in $T_{\rm eff}\sim200-1000$\,K and in log\,$g\sim0.04-0.1$.

Here we briefly review the laboratory astrophysics experiments to create radiation-driven hydrogen plasmas at WD photospheric conditions \citep{Falcon10b}.  These are aimed at measuring hydrogen spectral line shapes for the purpose of constraining theory and thus improving the spectroscopic method.  We then look to the future and discuss how we can utilize this experimental platform \citep{Falcon12b} to explore other outstanding problems in WD astronomy as well as in other areas of physics.

\section{(Brief) Review of the Experiments}

Please see \citet{Falcon12b} for a detailed description of our experimental platform.

\subsection{The {\it Z} Astrophysical Plasma Properties (ZAPP) Collaboration}

Ours is one of four laboratory astrophysics experiments being conducted in collaboration at the {\it Z} Pulsed Power Facility \citep{Matzen05} at Sandia National Laboratories.  The {\it Z} Astrophysical Plasma Properties (ZAPP) Collaboration \citep{Montgomery12} currently performs all these experiments simultaneously making use of the same x-ray source to drive plasma formation for a range of compositions.  ZAPP experimenters use these plasmas to study environments regarded as extreme on Earth but are quite normal with respect to the universe.  This includes reproducing the conditions at the solar convection zone boundary ($T_{\rm e}\sim 190$\,eV, $n_{\rm e}=10^{23}$\,cm$^{-3}$) for iron plasmas \citep{Bailey07,Bailey09}, producing highly photoionized ($T_{\rm e}\sim 20$--40\,eV, $n_{\rm i}\sim 10^{18}$\,cm$^{-3}$) neon plasmas \citep{Mancini09,Hall09,Hall10}, studying resonant Auger destruction \citep[e.g.,][]{Liedahl05} with photoionized silicon plasmas, and creating macroscopic hydrogen plasmas ($T_{\rm e}\sim 1$\,eV, $n_{\rm e}\sim 10^{17}$\,cm$^{-3}$) at WD photospheric conditions \citep{Falcon10b,Falcon12b}.

\subsection{Setup and Plasma Formation}

\begin{figure}[!h]
\begin{center}
\includegraphics[width=0.78\columnwidth]{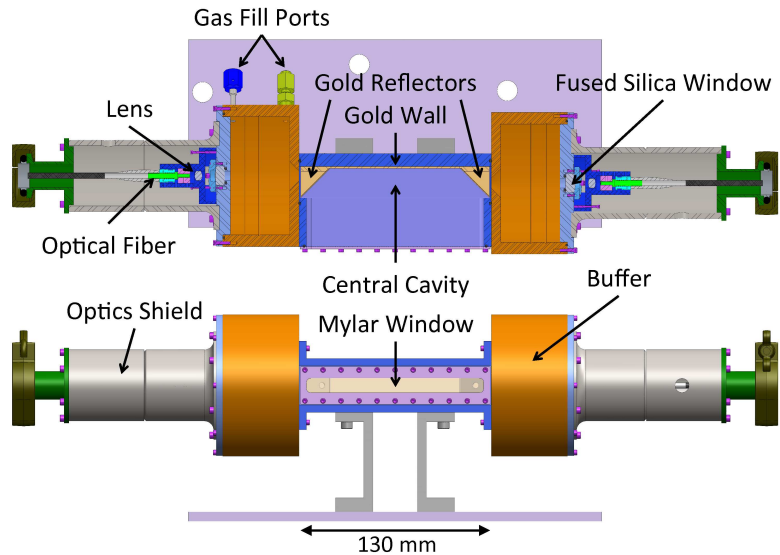}
\caption{Figure from \citet{Falcon12b}.  Top-view and front-view drawings of a gas cell design that uses two lines of sight at different distances from the gold wall and in antiparallel, horizontal directions.  The top-view is in cross-section, revealing gold reflectors used as back-lighting surfaces located on each end of the central cavity.  The buffers (orange) separate the optics from the hot plasma created in this cavity.
\label{drawing}}
\end{center}
\end{figure}

Before it becomes a plasma, we begin with a gas cell (Figure \ref{drawing}) filled with H$_2$ gas at room temperature.  We use a {\it z}-pinch dynamic hohlraum \citep{Sanford02,Rochau08} as the x-ray source to irradiate the gas cell.  X-rays transmit through a thin ($\sim 1.4$\,$\mu$m) Mylar window to enter the cell.  The Mylar attenuates the radiation, eliminating lower energy photons.  Hydrogen is transparent to photons at these higher energies; they stream through the gas and do not directly contribute to the plasma heating.  They are absorbed by a gold wall at the back end of the cell cavity.  The heated gold re-emits as a Planckian surface with a temperature of $\sim$5\,eV, heating the hydrogen through photoionization.  The radiation-driven quality of our plasma is unique amongst other plasma formation techniques.

\subsection{Data}

We observe the hydrogen plasma along lines of sight (LOS) perpendicular to the 
{\it z}-pinch radiation and parallel to the gold wall using lens-coupled optical fibers, which deliver the light to time-resolved spectrometer systems.

These systems consist of a 1\,m focal length, $f$/7 aperture Czerny-Turner design spectrometer with a streak camera that sweeps the spectrum over $\sim450$\,ns with $\sim1-2$\,ns temporal resolution.  The phosphor emittance from the camera goes through a micro-channel plate intensifier, and the output records onto Kodak T-MAX 400 film.  We achieve $\sim 9$\,\AA\ spectral resolution when using a 150\,groove/mm grating.

\begin{figure}[!h]
\begin{center}
\includegraphics[width=0.78\columnwidth]{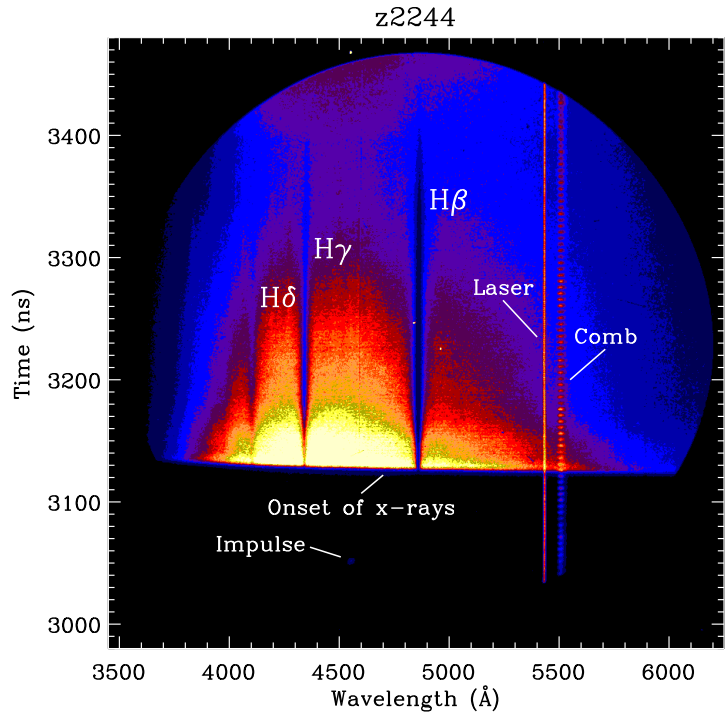}
\caption{Figure from \citet{Falcon12b}.  Time-resolved spectrum of hydrogen Balmer lines in absorption from experiment z2244.  The intensity of the back-lighting continuum decreases over the lifetime of the experiment as the gold reflector cools.
\label{absorption}}
\end{center}
\end{figure}

Figure \ref{absorption} shows an example time-resolved spectrum (streak image).  In the streak image, the hydrogen gas is invisible until the {\it z}-pinch x-rays arrive to heat the gold wall.  Within $\sim 80$\,ns from the onset of the x-rays, the plasma reaches a quasi-steady-state.  The line shapes remain relatively constant while the back-lighting continuum decreases in time as the gold reflector cools.

\section{Future Directions}

The motivation for these experiments and the experimental platform does not end at investigating hydrogen plasmas relevant to DA atmospheres.  A significant part of the scientific impetus is the potential to explore other interesting problems using the same general platform.  Here we mention some possible future directions for our experiments.  We focus on problems in WD astronomy, though the problems in other areas of physics beckon as well.

\subsection{Helium - DBs}

DBs show an increase in the mean and in the dispersion of surface gravity and hence spectroscopic mass below $T_{\rm eff}\approx 16,000$\,K \citep{Kepler07}.  Constraints from laboratory measurements on the line shape theory for helium lines at these plasma conditions \citep[e.g.,][]{Dimitrijevic90,Beauchamp97,BenChaouacha07} could strengthen the hypothesis that this large apparent mass dispersion might be real \citep{Bergeron11}.

Measuring gravitational redshifts in DB spectra using helium lines is possible \citep{Falcon12} but difficult due to the complication caused by pressure effects \citep{Greenstein67}.  More precise constraints on pressure shifts of individual He lines could lead to more DB mass determinations independent of the spectroscopic method.

In preliminary experiments performed in January 2011 using a He gas fill, we demonstrate that we can observe neutral helium lines in emission.  Understanding the He plasma formation using this platform, however, includes additional subtleties.  Here we expect ionizing photons to come from the {\it z}-pinch x-rays (attenuated through the Mylar window) directly as well as from the re-emitting gold wall.  Also, the first ionization potential of helium is at a higher energy than for hydrogen.

Immediate feasibility: moderate/high.

\subsection{Carbon/Oxygen - Hot DQs}

The first spectroscopic fits to observations of this relatively new class of WDs were poor \citep{Dufour08}, motivating new calculations of Stark broadened C~II line profiles \citep{Dufour11}.  The time for laboratory support for these calculations is now, as further work for O~I, O~II, and C~I line species is underway \citep{Dufour11}.

As in the He case, we expect ionizing photons to come from the attenuated {\it z}-pinch x-rays as well as from the re-emitting gold wall.  The first ionization potential of carbon is lower than for hydrogen, and for oxygen, it is similar.  The resulting ionization fraction should be higher for both species at a fixed gas fill pressure. 

Immediate feasibility: moderate/high.

\subsection{Magnetic Fields - DHs and DPs}

We will explore the feasibility of incorporating magnetic fields into our plasmas.  The {\it Z} Facility is capable of generating magnetic fields, but the difficulty of implementing this into our experimental platform in a controlled way pushes this project farther into the future.

Immediate feasiblity: low.

\section{Closing Remark}

The scientific potential and flexibility of our experimental platform is perhaps its greatest strength.  It can likely be used to address problems near the boundaries of our expertise and beyond, both in WD astronomy and in other areas of physics.  Because of this, here we solicit comments and scrutiny from you, the scientific community, so that we may optimize future experiments and investigations.

\acknowledgements This work was performed at Sandia National Laboratories.  We thank the {\it Z} dynamic hohlraum, accelerator, diagnostics, materials processing, target fabrication, and wire array teams, without which we cannot run our experiments.  Sandia is a multiprogram laboratory operated by Sandia Corporation, a Lockheed Martin Company, for the United States Department of Energy under contract DE-AC04-94AL85000.  This work has made use of NASA's Astrophysics Data System Bibliographic Services.  R.E.F. acknowledges support of the National Physical Science Consortium, and R.E.F., M.H.M., and D.E.W. gratefully acknowledge support of the Norman Hackerman Advanced Research Program under grant 003658-0252-2009.


\bibliographystyle{asp2010}
\bibliography{falcon}

\end{document}